\documentclass[
aps,%
12pt,%
final,%
notitlepage,%
oneside,%
onecolumn,%
nobibnotes,%
nofootinbib,% 
superscriptaddress,%
noshowpacs,%
fromemail=on,%
centertags]%
{revtex4-2}

\usepackage{graphicx}
\usepackage{subcaption}

\begin{document}

%\selectlanguage{english}

\title{Muon Track Reconstruction Procedures \\
at the Baikal-GVD Neutrino Telescope\\
}% Разбиение на строки осуществляется командой \\

\author {V. A.~Allakhverdyan}
\affiliation{ Joint Institute for Nuclear Research, Dubna, Russia}

\author{A. D.~Avrorin}
\affiliation{ Institute for Nuclear Research of the Russian Academy of Sciences, 
%60th October Anniversary Prospect 7a, 
Moscow, Russia}

\author{A. V.~Avrorin}
\affiliation{ Institute for Nuclear Research of the Russian Academy of Sciences, 
%60th October Anniversary Prospect 7a, 
Moscow, Russia}

\author{V. ~M.~ Aynutdinov}
\affiliation{ Institute for Nuclear Research of the Russian Academy of Sciences, 
%60th October Anniversary Prospect 7a, 
Moscow, Russia}

\author{Z. ~Barda\v{c}ov\'{a}}
\affiliation{ Comenius University, Bratislava, Slovakia}
\affiliation{Czech Technical University, Institute of Experimental and Applied Physics, CZ-11000 Prague, Czech Republic}

\author{I. A.~Belolaptikov}
\affiliation{ Joint Institute for Nuclear Research, Dubna, Russia}

\author{E. ~A.~ Bondarev}
\affiliation{ Institute for Nuclear Research of the Russian Academy of Sciences, 
%60th October nniversary Prospect 7a, 
Moscow, Russia}

\author{I. V.~Borina}
\affiliation{ Joint Institute for Nuclear Research, Dubna, Russia}

\author{N. M.~Budnev}
\affiliation{ Irkutsk State University, Irkutsk, Russia}

\author{V. A.~Chadymov}
\affiliation{ Independed researcher}

\author{A. S.~Chepurnov}
\affiliation{ Skobeltsyn Research Institute of Nuclear Physics, Moscow State University, Moscow, Russia}

\author{V. Y.~Dik}
\affiliation{ Joint Institute for Nuclear Research, Dubna, Russia}
\affiliation{Institute of Nuclear Physics ME RK, Almaty, Kazakhstan}

\author{\fbox{G. V.~Domogatsky}}
\affiliation{ Institute for Nuclear Research of the Russian Academy of Sciences, 
%60th October Anniversary Prospect 7a, 
Moscow, Russia}

\author{A. A.~Doroshenko}
\affiliation{ Institute for Nuclear Research of the Russian Academy of Sciences, 
%60th October Anniversary Prospect 7a, 
Moscow, Russia}

\author{R.~Dvornick\'{y}}
\affiliation{ Comenius University, Bratislava, Slovakia}
\affiliation{Czech Technical University, Institute of Experimental and Applied Physics, CZ-11000 Prague, Czech Republic}

\author{A. N.~Dyachok}
\affiliation{ Irkutsk State University, Irkutsk, Russia}

\author{Zh.-A. M.~Dzhilkibaev}%
\affiliation{ Institute for Nuclear Research of the Russian Academy of Sciences, 
%60th October Anniversary Prospect 7a, 
Moscow, Russia}

\author{E. ~Eckerov\'{a}}
\affiliation{ Comenius University, Bratislava, Slovakia}
\affiliation{Czech Technical University, Institute of Experimental and Applied Physics, CZ-11000 Prague, Czech Republic}

\author{T. V.~Elzhov}
\affiliation{ Joint Institute for Nuclear Research, Dubna, Russia}

\author{V. N.~Fomin}
\affiliation{ Independed researcher}

\author{A. R.~Gafarov}
\affiliation{ Irkutsk State University, Irkutsk, Russia}

\author{K. V.~Golubkov}
\affiliation{ Institute for Nuclear Research of the Russian Academy of Sciences, 
%60th October Anniversary Prospect 7a, 
Moscow, Russia}

\author{N. S.~Gorshkov}
\affiliation{ Joint Institute for Nuclear Research, Dubna, Russia}

\author{T. I.~Gress}
\affiliation{ Irkutsk State University, Irkutsk, Russia}

\author{K. G.~Kebkal}
\affiliation{ LATENA, St. Petersburg, Russia}

\author{V. K.~Kebkal}
\affiliation{ LATENA, St. Petersburg, Russia}

\author{I. V.~Kharuk}
\affiliation{ Institute for Nuclear Research of the Russian Academy of Sciences, 
%60th October Anniversary Prospect 7a, 
Moscow, Russia}

\author{E. V.~Khramov}
\affiliation{ Joint Institute for Nuclear Research, Dubna, Russia}

\author{M. M.~Kolbin}
\affiliation{ Joint Institute for Nuclear Research, Dubna, Russia}

\author{S. O.~Koligaev}
\affiliation{ INFRAD, Dubna, Russia}

\author{K. V.~Konischev}
\affiliation{ Joint Institute for Nuclear Research, Dubna, Russia}

\author{A. V.~Korobchenko}
\affiliation{ Joint Institute for Nuclear Research, Dubna, Russia}

\author{A. P.~Koshechkin}
\affiliation{ Institute for Nuclear Research of the Russian Academy of Sciences, 
%60th October Anniversary Prospect 7a, 
Moscow, Russia}

\author{V. A.~Kozhin}
\affiliation{ Skobeltsyn Research Institute of Nuclear Physics, Moscow State University, Moscow, Russia}

\author{M. V.~Kruglov}
\affiliation{ Joint Institute for Nuclear Research, Dubna, Russia}

\author{V. F.~Kulepov}
\affiliation{ Nizhny Novgorod State Technical University, Nizhny Novgorod, Russia}

\author{A. A.~Kulikov}
\affiliation{ Irkutsk State University, Irkutsk, Russia}

\author{Y. E.~Lemeshev}
\affiliation{ Irkutsk State University, Irkutsk, Russia}

\author{R. R.~Mirgazov}
\affiliation{ Irkutsk State University, Irkutsk, Russia}

\author{D. V.~Naumov}
\affiliation{ Joint Institute for Nuclear Research, Dubna, Russia}

\author{A. S.~Nikolaev}
\affiliation{ Skobeltsyn Research Institute of Nuclear Physics, Moscow State University, Moscow, Russia}

\author{I. A.~Perevalova}
\affiliation{ Irkutsk State University, Irkutsk, Russia}

\author{D. P.~Petukhov}
\affiliation{ Institute for Nuclear Research of the Russian Academy of Sciences, 
%60th October Anniversary Prospect 7a, 
Moscow, Russia}

\author{E. N.~Pliskovsky}
\affiliation{ Joint Institute for Nuclear Research, Dubna, Russia}

\author{M. I.~Rozanov}
\affiliation{ St. Petersburg State Marine Technical University, St. Petersburg, Russia}

\author{E. V.~Ryabov}
\affiliation{ Irkutsk State University, Irkutsk, Russia}

\author{G. B.~Safronov}
\email{E-mail: grigorybs@gmail.com}
\affiliation{ Institute for Nuclear Research of the Russian Academy of Sciences, 
%60th October Anniversary Prospect 7a, 
Moscow, Russia}

\author{B. A.~Shaybonov}
\affiliation{ Joint Institute for Nuclear Research, Dubna, Russia}

\author{E. V.~Shirokov}
\affiliation{ Skobeltsyn Research Institute of Nuclear Physics, Moscow State University, Moscow, Russia}

\author{V. Y.~Shishkin}
\affiliation{ Skobeltsyn Research Institute of Nuclear Physics, Moscow State University, Moscow, Russia}

\author{F.~\v{S}imkovic}
\affiliation{ Comenius University, Bratislava, Slovakia}
\affiliation{Czech Technical University, Institute of Experimental and Applied Physics, CZ-11000 Prague, Czech Republic}

\author{A. E. Sirenko}
\affiliation{ Joint Institute for Nuclear Research, Dubna, Russia}

\author{A. V.~Skurikhin}
\affiliation{ Skobeltsyn Research Institute of Nuclear Physics, Moscow State University, Moscow, Russia}

\author{A. G.~Solovjev}
\affiliation{ Joint Institute for Nuclear Research, Dubna, Russia}

\author{M. N.~Sorokovikov}
\affiliation{ Joint Institute for Nuclear Research, Dubna, Russia}

\author{I.~\v{S}tekl}
\affiliation{Czech Technical University, Institute of Experimental and Applied Physics, CZ-11000 Prague, Czech Republic}

\author{A. P.~Stromakov}
\affiliation{ Institute for Nuclear Research of the Russian Academy of Sciences, 
%60th October Anniversary Prospect 7a, 
Moscow, Russia}

\author{O. V.~Suvorova}
\affiliation{ Institute for Nuclear Research of the Russian Academy of Sciences, 
%60th October Anniversary Prospect 7a, 
Moscow, Russia}

\author{V. A.~Tabolenko}
\affiliation{ Irkutsk State University, Irkutsk, Russia}

\author{V. I.~Tretjak}
\affiliation{ Joint Institute for Nuclear Research, Dubna, Russia}

\author{B. B.~Ulzutuev}
\affiliation{ Joint Institute for Nuclear Research, Dubna, Russia}

\author{Y. V.~Yablokova}
\affiliation{ Joint Institute for Nuclear Research, Dubna, Russia}

\author{D. N.~Zaborov}
\affiliation{ Institute for Nuclear Research of the Russian Academy of Sciences, 
Moscow, Russia} 

\author{S. I.~Zavyalov}
\affiliation{ Joint Institute for Nuclear Research, Dubna, Russia}

\author{D. Y.~Zvezdov}
\affiliation{ Joint Institute for Nuclear Research, Dubna, Russia}
%\collaboration{59}{(Baikal-GVD Collaboration)}
%%\affiliation{}

\collaboration{Baikal-GVD Collaboration}%\noaffiliation

\date{Received August 16, 2025; revised August 16, 2025; accepted August 16, 2025}
%\today печатает cегодняшнее число

\begin{abstract}
The Baikal-GVD neutrino telescope is the largest neutrino detector of its kind in the Northern Hemisphere. Muons produced in neutrino interaction in the vicinity of the detector leave track-like response in the detector allowing to reconstruct the neutrino arrival direction with the precision up to $0.2^{\circ}$. The Baikal-GVD collaboration has developed a variety of methods for the track-like event analysis. Methods for track-like event direction and energy reconstruction and neutrino cadidate event selection are discussed in this report. Preliminary results of application of analysis pipeline to the data-taking seasons from 2019 to 2021 are shown.
\end{abstract}

\maketitle

\section{Introduction}

The main purpose of the large-volume neutrino telescopes such as Baikal-GVD is the study of TeV -- PeV cosmic neutrino flux.  
The Baikal-GVD telescope being constructed in Lake Baikal presently consists of 14 independent sub-detectors - clusters, carrying in total 4212 optical modules enabling the detection volume in high-energy cascade detection channel of $>$0.6 km$^3$. The data from the partially completed detector has allowed to obtain a number of results such as observation of atmospheric neutrino flux in track channel \cite{1} and a measurement of the diffuse astrophysical neutrino flux using cascade-like events \cite{2}. While the cascade detection channel provides precise neutrino energy reconstruction and is good for spectrum measurements, the track channel is the best for the purpose of the astrophysical point source searches thanks to better direction reconstruction. The pointing accuracy in track channel reaches $\sim0.2^{\circ}$ and better for sufficiently long tracks. In these proceedings we discuss the status of the track-like events analysis pipeline in Baikal-GVD, present methods adopted for track-like event reconstruction and selection. Discuss the results of application of these methods to data collected in seasons from 2019 to 2021. 

In Section 2 details of the detector setup and the data flow are given. In Section 3 the muon track reconstruction methods are described in particular lake noise suppression, track reconstruction and muon energy estimation algorithms. In section 3 the track-like neutrino event selection methods are described. Finally in Section 4 the results of application of these methods to data collected in seasons from 2019 to 2021 are demonstrated.

%The measured spectrum of cosmic ray (CR) protons extend to beyond $10^{21}$eV - far beyond energies accessible at man-made accelerators such as LHC. This proves an existence of powerful cosmic systems accelerating charged particles to such energies. Neutrinos produced in proton interactions near the acceleration site can sched light on the physics of CR accelerating systems. Indeed neutrinos propagate at cosmological distances without scattering or absorption not affected by cosmic magnetic fields. Once detected neutrinos can be associated to their source and provide the direct probe of energy scale in its vicinity. 

\section{The Baikal-GVD neutrino telescope}

The Baikal-GVD neutrino telescope in its present configuration consists of 14 independent sub-detectors -- clusters (Fig. 1). The detector volume of $>$0.6 km$^3$ makes it the largest neutrino telescope in the Northern Hemisphere. The lake depth at the telescope location is about 1366 meters, water light absorption length reaches 22 m. Each cluster includes 8 or 9 strings instrumented with 36 optical modules (OM) at depths between 750 and 1275 meters with a vertical spacing of 15 m. An OM is a hermetic glass container instrumented with high-quantum-efficiency 10-inch PMT HAMAMATSU R7081-100 and various sensors. Each string is subdivided onto three elementary units of the data readout -- sections. Each section consists of 12 OMs each connected via coaxial cable to the section central module (CM). CM receives analog signal from OMs, digitizes it and performs triggering. The trigger condition within section is satisfied when there are at least two PMT pulses at nearby OMs within the time window of 100 ns with amplitudes at least above $A_{\text{high}}$ and $A_{\text{low}}$, where $A_{\text{high}}$ and $A_{\text{low}}$ are trigger thresholds which are adjusted according to background noise conditions and may vary across the detector. On average $A_{\text{high}}=4.5$ and $A_{\text{low}}=1.5$ photoelectrons. Once the trigger signal from any CM of the cluster is received by the cluster center (CC) the CC requests the buffer readout from each CM of the cluster. Upon recieval of such request each CM sends 5 $\mu$s time window to the shore station. The data is further immediately transmitted to JINR, Dubna.  At JINR the primary RAW data processing is performed during which independent readout windows from the CMs are merged into the cluster-wide events \cite{3}. Further processing is performed via two pipelines: multi-cluster and single-cluster. In multi-cluster pipeline common events between clusters are found. Single-cluster events are merged into the multi-cluster events based on requirement that time of the triggers in different clusters is consistent with the response from the same particles. In further processing time and charge calibrations are applied to single-cluster or multi-cluster events \cite{4,5}. OM coordinates, as reconstructed in quasi-online manner from regular acoustic modem polls \cite{6}, are available from the online database. In present work we discuss results of data processing in track-like single-cluster pipeline while the multi-cluster track-like analysis pipeline is still in development.

\section{Muon track reconstruction}

The dominant source of PMT pulses (or hits) in the telescope data is the ambient optical background of the Lake Baikal waters. The backround is due to luminiscense in layers of sinking remains of living organisms and/or algae. The PMT signal due to background is at the 1 photoelectron level. The rate is not uniform and depends on the depth and season reaching up to hundreds of kHz at top OM layers in mid-summer while in general being few 10's of kHz per detection channel \cite{7}. Given the 5 $\mu$s event time window, the number of noise hits per the single-cluster event may reach up to hundred and beyond. 

\subsection{Muon Hit Finding and Direction Reconstruction}

At the first step of the muon reconstruction procedure PMT hits due to Cerenkov light are identified while those due to lake optical background are suppressed. This is done by means of the two-step criteria employing Cerenkov pulses correlation in time. At the first step the simple criterium is used, difference of times of hits produced by the Cerenkov should satisfy the following condition: $\Delta t < \Delta R n/c \, + \, \delta$, where $\Delta R$ is the distance between modules, $c/n$ is the speed of light in water and $\delta$ incorporates fluctuations due to time measurement precision and is typically set to 10 ns. In the event analysis process the collection of hits forming the largest fully-connected clique is found using graph theory methods such as Bron\textendash Kerbosch algorithm \cite{8}. 
The preliminary track direction estimation is obtained after this step using the sum of vectors formed by time-ordered hits, in an assumption of flat wavefront propagating through the detector.
At the second step the so-called directional causality criterium is used which is described in detail elsewhere \cite{8}, in this criterium hits are connected using strict requirements depending on the track direction. The scan on uniform rectangular grid in $(\theta,\phi)$ coordinates defined in the $60^{\circ}{\times}60^{\circ}$ region around preliminary track direction is performed. Track directions providing the set of largest fully-connected hit cliques are identified. For each such direction the fit of the track position is performed using the m-estimator loss function \cite{8}. Few hit collections giving best m-estimator values are passed to the full track reconstruction which is seeded by respective track directions and positions. The hit finding procedure provides the purity of hit collection of about 95\% keeping the probability of capturing the Cerenkov hit at the level of 95\% on average for atmospheric neutrino spectrum \cite{8}.

The track direction and position reconstruction is performed by minimising the two term loss function:

\begin{equation}
Q({\bf v},{\bf x},t) = \sum_{i} ({(t_{i}-t_{\text{th}}({\bf v},{\bf x},t))^2 \over \sigma^2} + w(N_{\text{hits}},q_{\text{sum}})f(q_{i})g(r_{i}))
\end{equation}

%f(N_{hits},q_{sum},r_{i},q_{i}))
where the first term is the $\chi2$ based on $t_{\text{th}}$ calculated in assumption of direct Cerenkov light emission from the muon, $\sigma$ is the time measurement precision estimate set to the value of 3 ns and $t_i$ is the measured hit time. The second term introduces the penalty for offset tracks based on the distance from the track and deposited charge corrected for the OM angular acceptance. Charge and distance functions $f(q_{i})$ and $g(r_{i})$ are described in detail elsewhere \cite{1} while $w(N_{\text{hits}},q_{\text{sum}})$ is the weight of second term adjusted to maintain its contribution at similar level for the wide muon energy range from TeV up to PeV and beyond. 
%Here the $f(q_{i})$ and $g(r_{i})$ are functions of hit charge corrected for the angular sensitivity and distance from the track respectively and $w(N_{hits},q_{sum})$ is the weight of the second term introduced as a function of the number of hits and total charge deposition from the track.
% The weight of the charge term is optimised to maintain $\sim$similar contribution of charge term into the loss function over the wide energy range.

%For each of the few received hit collections the minimisation of the loss function with successive of hits one-by-one to account for possible remaining noise hits is performed. 
For each of the few received hit collections the minimization of the loss function with successive variations of hit collections to account for possible remaining noise hits is performed.
The hit configuration giving the best loss function value is used for the muon energy estimation and calculation of variety of track quality parameters to be used at the stage of the event selection. 
The median angular resolution of $0.2^{\circ}$ is achieved for sufficiently long tracks for the described track reconstruction procedure (Fig. 2).   

% In the causality criterium it is required that time between any two hits is smaller than the light propagation time between respective OMs: $\Deltat < \DeltaR n/c + \delta$ where $\delta$ is typically 10ns. Directional causality criterium puts    

\subsection{Muon Energy Reconstruction}

High-energy muons crossing the detector sensitive volume loose energy mainly via the bremsstrahlung and pair production and above $\sim$1 TeV the rate of losses is linearly proportional to the muon energy. That makes an energy loss rate estimate a viable proxy for the muon energy reconstruction. An energy loss rate is estimated by means of the so-called median energy estimator. In this approach the amount of light emitted by the muon is estimated for each OM within 40 m from the track based on charge deposition and taking into account light absorption, geometric factor and OM angular sensitivity. Such elementary energy losses are ordered by their magnitude and the median value, $k$, is taken as a proxy. The proxy is mapped to a muon energy $E^{true}_{\mu}$ in the plane containing the geometric center of the detector via the polynomial fit of median of the $E^{\text{true}}_{\mu}$ distribution in bins of $k$ as a function of $k$. The 68\% confidence interval for the muon energy is found using polynomial fits of respective quantiles of $E^{\text{true}}_{\mu}$ distribution in bins of $k$. The resulting estimate of $E^{\text{true}}_{\mu}$ experience up to 20\% bias vs number of hits in the event and zenith angle for the $\Phi\sim E^{-2}$ neutrino spectrum. Respective correction factors were derived for this spectrum and applied to $E^{\text{true}}_{\mu}$ quantile estimates. Resulting bias and uncertainty of the muon energy measurement are shown at Fig. 3. The energy resolution of factor 2.5 is attained for muons of energy above $\sim$10 TeV.

\section{Neutrino candidate event selection}

The described track reconstruction procedure was applied to the data and MC of data-taking seasons 2019\textendash 2021 in the single cluster-regime. Reconstructed events are dominated by muon bundles and the total rate of events passing the reconstruction procedure is $\sim$3 Hz per cluster. In this analysis the region of interest for the track-like neutrino events are upgoing events with zenith angles ${>}100^{\circ}$, the zenith angle distribution for such events is shown at Fig. 4. The data is compared to the expectations from various MC-simulated events passed through the realistic detector simulation for each season including realistic trigger thresholds, the lake noise level, missing channels due to hardware failures and other detector-related parameters. The atmospheric muon bundle MC is based on the bank of CORSIKA 7.741 \cite{9} events. Atmospheric neutrino flux is simulated according to the Bartol flux model \cite{10}. Finally the astrophysical neutrino flux corresponds to the IceCube measurement of $\nu_{\mu}$ flux \cite{11}. Neutrino fluxes take into account neutrino absorption in Earth evaluated with nuFATE program \cite{12}. In addition a dedicated bank of CORSIKA events with the energy of leading muon ${>}100$ TeV was used to enrich the background MC simulation with a rare very-high-energy muon bundle events. The good data-MC agreement is demonstrated for the rate of misreconstructed atmospheric muon bundle events which exceeds the rate of atmospheric neutrino by the factor of 10$^2$\textendash 10$^4$ depending on the zenith angle. Misreconstructed muon events are rejected with the widely used machine learning technique -- boosted decision trees (BDTs). Such BDTs use a list of variables obtained in the event reconstruction process and return a floating point number quantifying the degree of belief that event is background or signal. 

Two BDTs aimed at low-energy and high-energy neutrino selection were constructed using slightly different lists of $\sim$20 quasi-independent variables incorporating the fit quality and the event topology, e.g. fitting function (1) divided by the number of degrees of freedom, track length, fraction of hits within 22 m from the track and others. BDTs were trained using different signal event spectra for the reconstructed zenith angle region $\theta > 100^{\circ}$. The low-energy BDT (BDT\_LE) was trained on atmospheric neutrino spectrum with cutoff of $E_{\mu}<10$ TeV. The high-energy BDT (BDT\_HE) was trained on astrophysical spectrum $\Phi\sim E^{-2}$ with an energy cut $E_{\mu}>10$ TeV. Two CORSIKA-based muon bundle samples were used as a background -- regular spectrum and events with leading muons with $E>100$ TeV. The TMVA framework was used for the BDT construction and training \cite{13}. BDTs were constructed using the AdaBoost boosting technique. 

The BDT response distributions are shown at Fig. 5, events due to atmospheric background are concentrated in the left part of the distribution for each BDT while neutrino events concentrate in the right part. The cut at the classifier response of BDT\_LE $>$ 0.25 OR BDT\_HE $>$ 0.25 provides the sample of events with misreconstructed muon bundle contribution estimated as $\sim$3\% and neutrino event efficiency being of order of 70\%. The cut value was chosen to maintain specific muon bundle background contamination and was not optimized for any specific analysis. 

\section{Results of the data processing}

An event selection procedure was applied to the data sample of data-taking seasons 2019\textendash 2021. The total livetime of the sample corresponds to the 14.4 years of data -taking in single-cluster regime. An event sample of 1189 neutrino candidates was selected with the aforementioned cuts on the BDT classifiers response. An event rate in data was found to be ~30\% larger than expected from the simulation of atmospheric neutrino using the Bartol flux, work is underway to determine the reasons for the observed discrepancy. The normalization of MC for atmospheric neutrino and astrophysical neutrino was scaled by the single constant factor 1.32 for further data-MC comparison. Distributions of number of hits used for the reconstruction and reconstructed muon energy are shown at Fig. 6. An acceptable agreement of the data and MC expectation is demonstrated. 

\section{Summary}
Methods and procedures for the track-like events reconstruction at the Baikal-GVD neutrino telescope were discussed. Reconstruction algorithms allows to reconstruct the muon direction with the precision up to $0.2^{\circ}$ for track of $>$500 m length. The muon energy is reconstructed with the precision of the factor 2.5. Misreconstructed atmospheric muon background is suppressed with the boosted decision trees optimized for efficient neutrino selection in TeV\textendash PeV energy range. The analysis pipeline was applied to data-taking seasons 2019\textendash 2021 with no further cut optimization. The demonstration sample of 1189 neutrino candidates was selected. Data-MC agreement for the neutrino candidate events was verified on the demonstration sample of events.

This work was supported by Russian Science Foundation grant No. 24-72-10056. The research results were obtained using the material and technical base of the Baikal Neutrino Observatory and experimental data accumulated by the Baikal-GVD deep-sea neutrino telescope.

%
% Список литературы
%

%

\newpage
\textbf{Fig. 1.} The Baikal-GVD neutrino telescope in the configuration deployed in February - March 2025.

\textbf{Fig. 2.} Precision of muon direction reconstruction as a function of track length.

\textbf{Fig. 3.} Muon energy measurement bias and uncertainty as a function of reconstructed energy (a) and number of hits (b).

\textbf{Fig. 4.} Angular distribution of tracks for the region of zenith angles $\theta > 100^{\circ}$ reconstructed in the data of seasons from 2019 to 2021 before the neutrino selection cuts. The data is compared to the expectation from the CORSIKA -based muon bundle MC (red), muon neutrino MC based on Bartol flux (blue) \cite{10} and expectation from astrophysical neutrino flux based on IceCube $\nu_{\mu}$ spectrum fit (green) \cite{11}. In addition contribution from the sample of events based on CORSIKA with the leading muon $E > 100$ TeV is shown (magenta). The event rate is dominated by the misreconstructed muon bundle events which are further suppressed with the neutrino candidate selection cuts.

\textbf{Fig. 5.} Distributions of boosted decision tree classifiers response for low-energy BDT classifier (a) and high-energy classifier (b) for events reconstructed in the data from seasons 2019-2021 compared to signal and background MC expectations.

\textbf{Fig. 6.} Distributions of number of track hits (a) and reconstructed muon energy (b) obtained in the data from seasons 2019-2021 compared to the expectations from MC.

\newpage
\begin{figure}
%\begin{center}
	\includegraphics[width=12.4 cm]{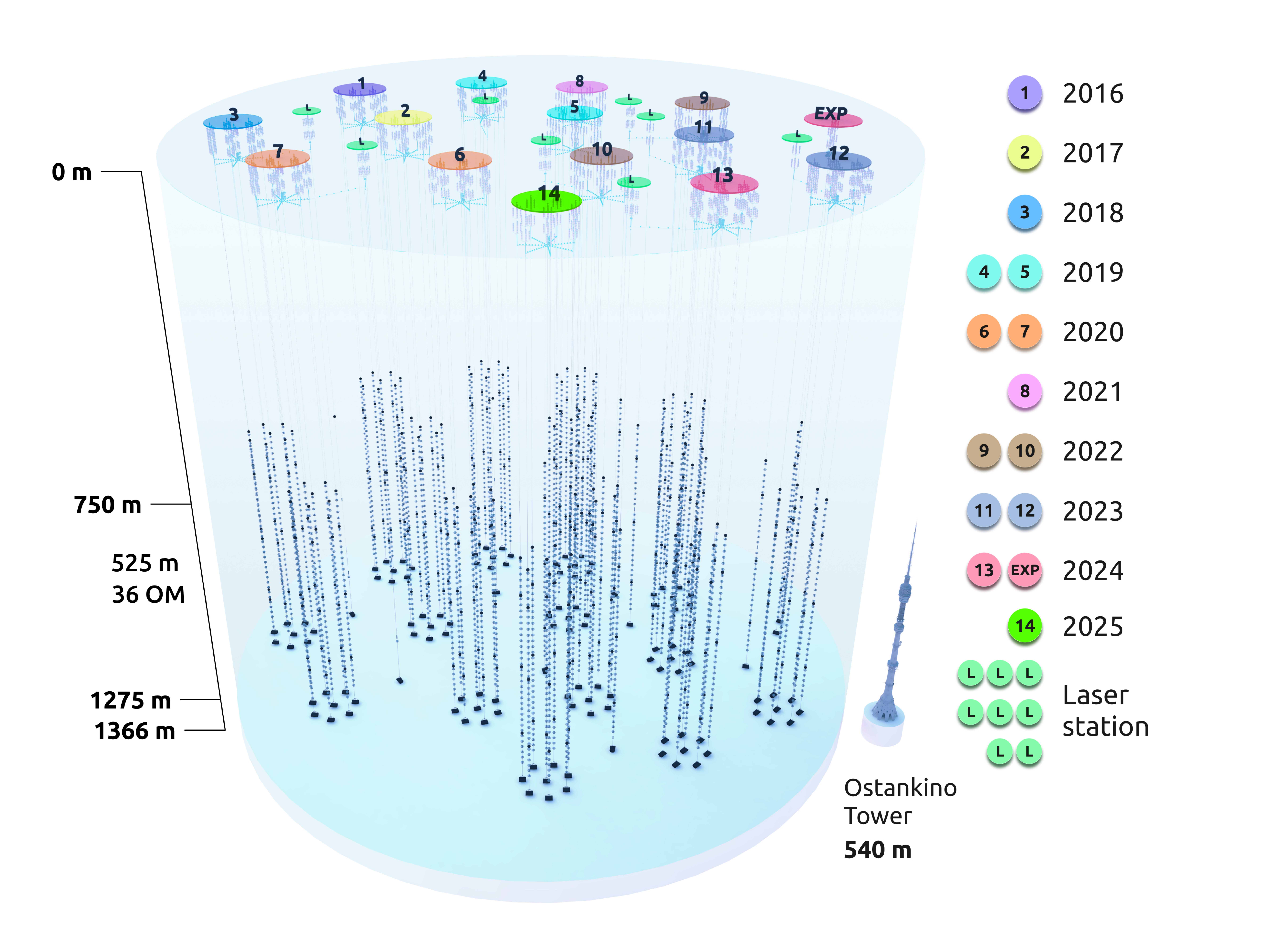}
	\caption{The Baikal-GVD neutrino telescope in the configuration deployed in February - March 2025.}
%\end{center}
\end{figure}

\begin{figure}
\begin{center}
	\includegraphics[width=8.5 cm]{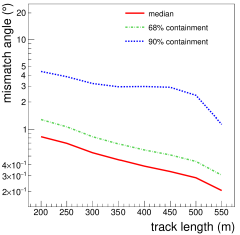}
	\caption{Precision of muon direction reconstruction as a function of track length.}
\end{center}
\end{figure}

\begin{figure}
\begin{center}
	\includegraphics[width=14.5 cm]{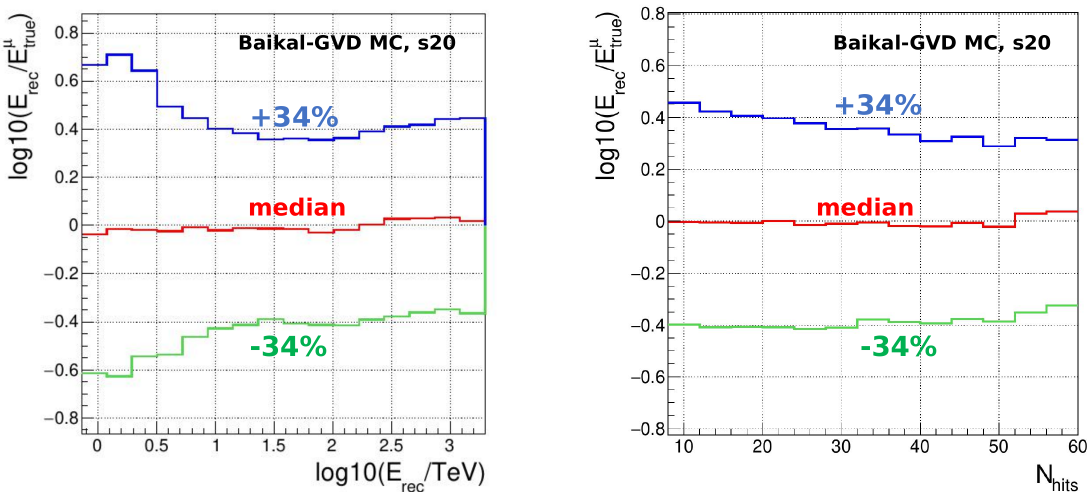}
	\caption{Muon energy measurement bias and uncertainty as a function of reconstructed energy (a) and number of hits (b).}
\end{center}
\end{figure}

\begin{figure}
\begin{center}
	\includegraphics[width=8.5 cm]{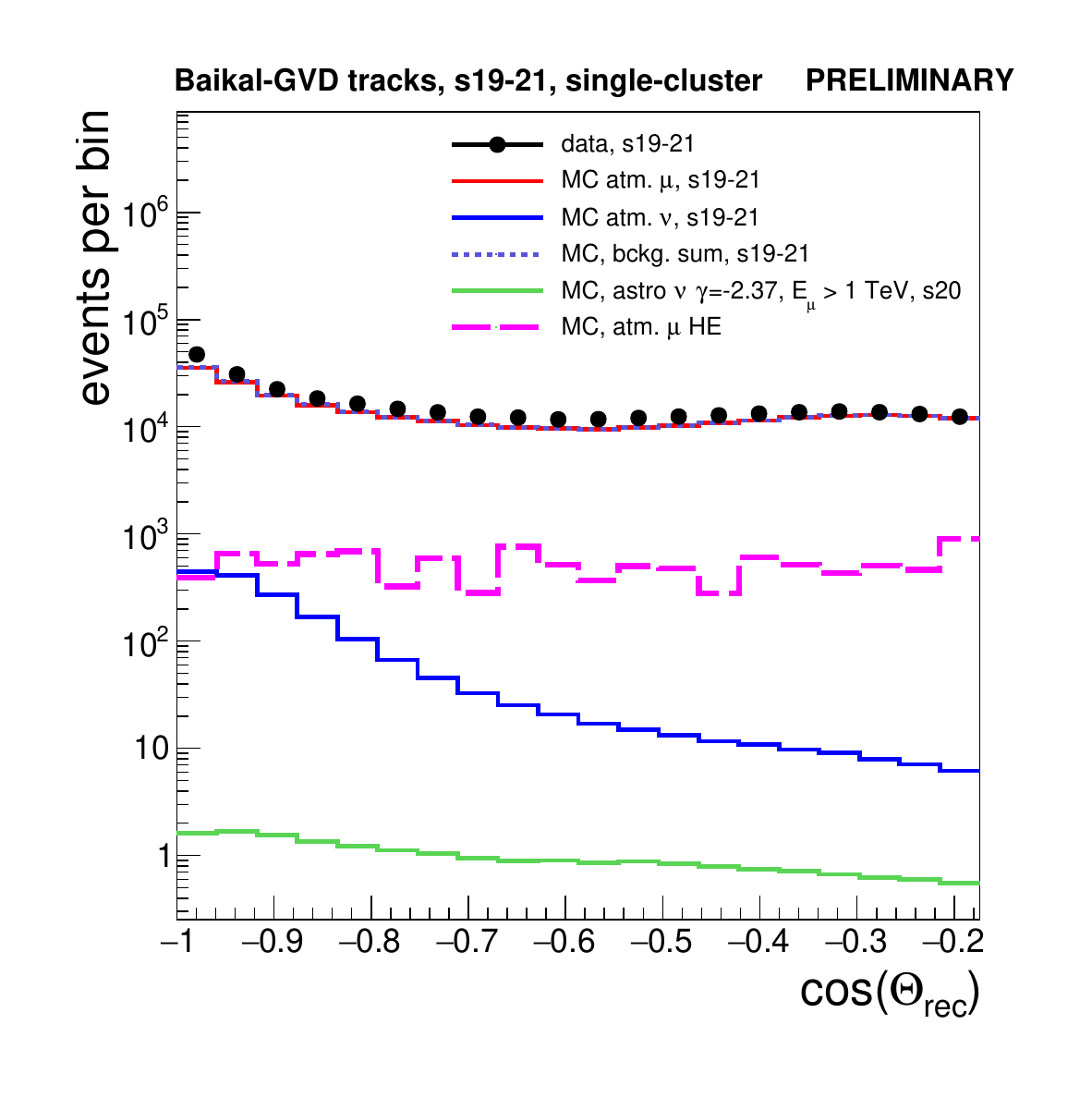}
	\caption{Angular distribution of tracks for the region of zenith angles $\theta > 100^{\circ}$ reconstructed in the data of seasons from 2019 to 2021 before the neutrino selection cuts. The data is compared to the expectation from the CORSIKA -based muon bundle MC (red), muon neutrino MC based on Bartol flux (blue) \cite{10} and expectation from astrophysical neutrino flux based on IceCube $\nu_{\mu}$ spectrum fit (green) \cite{11}. In addition contribution from the sample of events based on CORSIKA with the leading muon $E > 100$ TeV is shown (magenta). The event rate is dominated by the misreconstructed muon bundle events which are further suppressed with the neutrino candidate selection cuts.}
\end{center}
\end{figure}

\begin{figure}
\begin{center}
	\begin{subfigure}{0.49\textwidth}
	\includegraphics[width=7.5 cm]{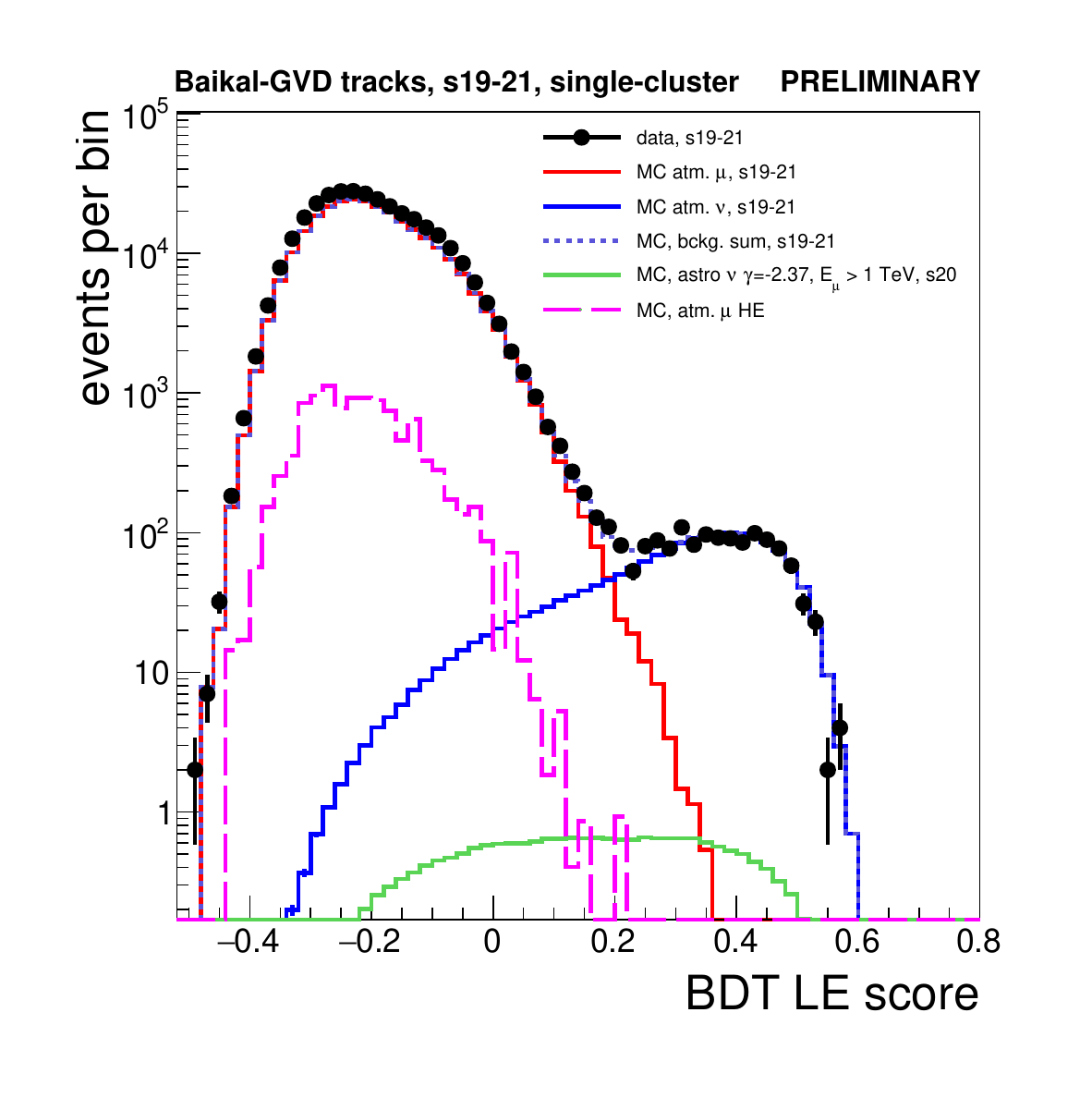}
	%\caption{}
	\end{subfigure}
	\begin{subfigure}{0.49\textwidth}
	\includegraphics[width=7.5 cm]{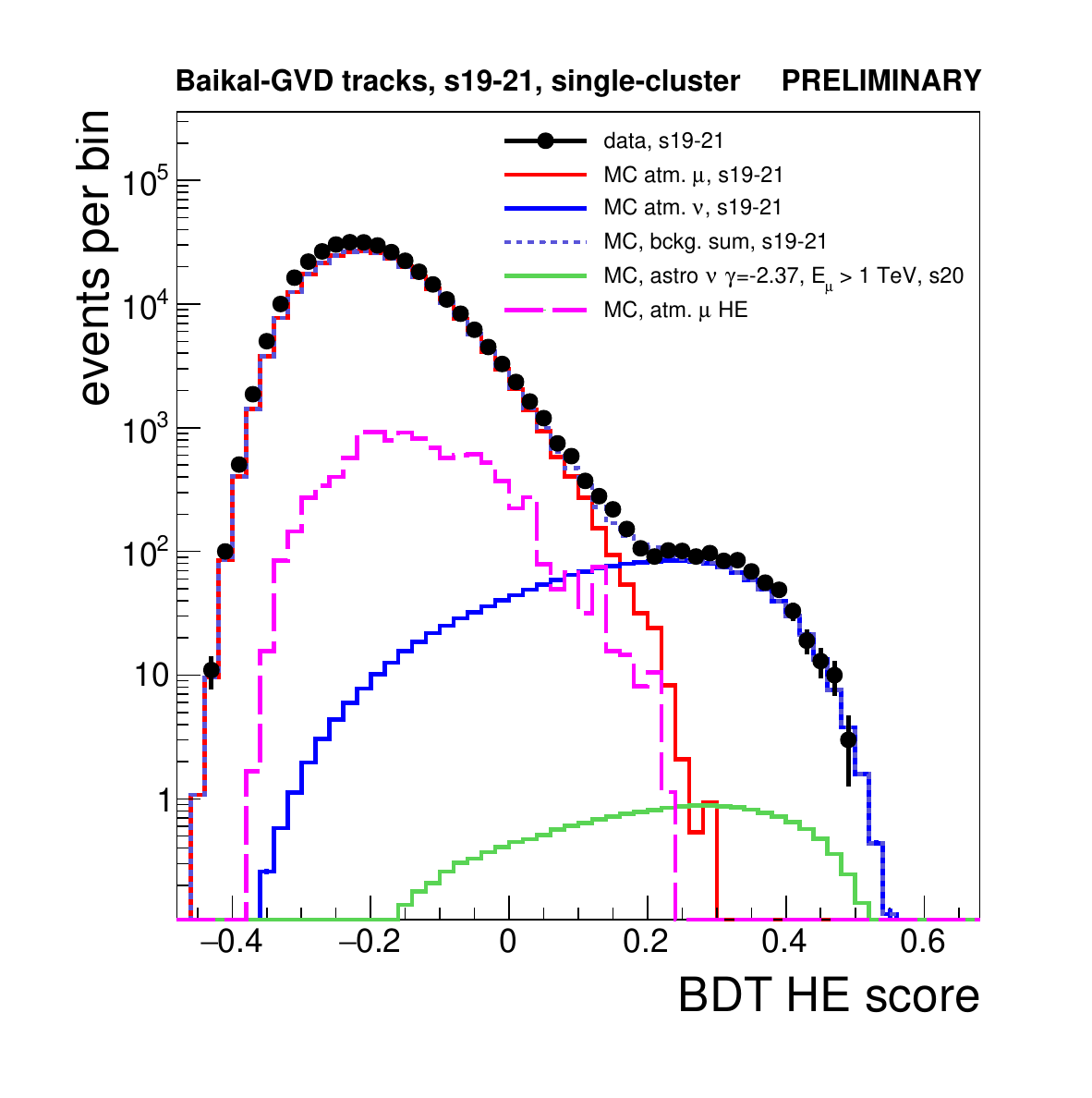}
	\end{subfigure}
	\caption{Distributions of boosted decision tree classifiers response for low-energy BDT classifier (a) and high-energy classifier (b) for events reconstructed in the data from seasons 2019-2021 compared to signal and background MC expectations.}
\end{center}
\end{figure}

\begin{figure}
\begin{center}
	\begin{subfigure}{0.49\textwidth}
	\includegraphics[width=7.5 cm]{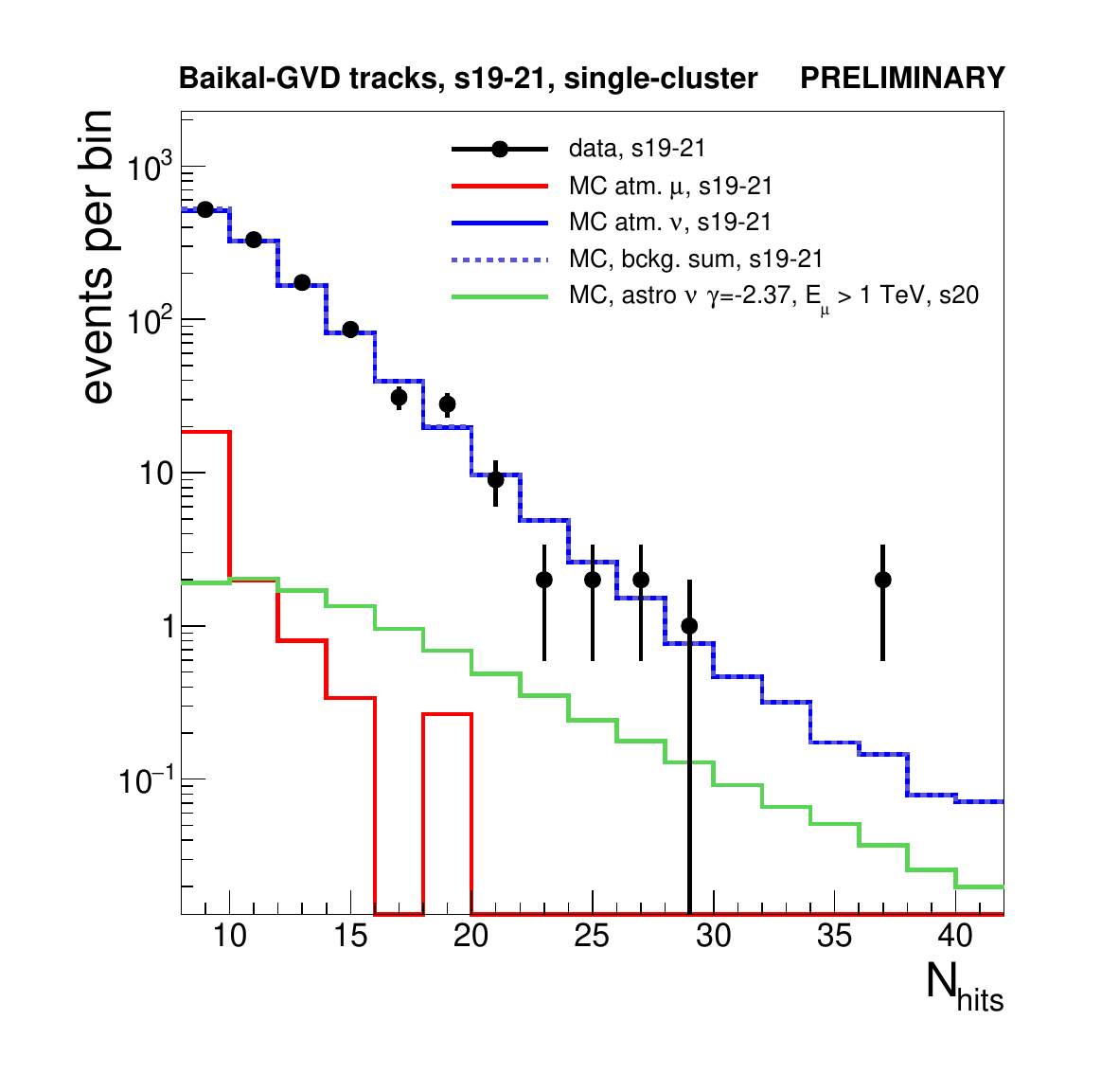}
	%\caption{}
	\end{subfigure}
	\begin{subfigure}{0.49\textwidth}
	\includegraphics[width=7.5 cm]{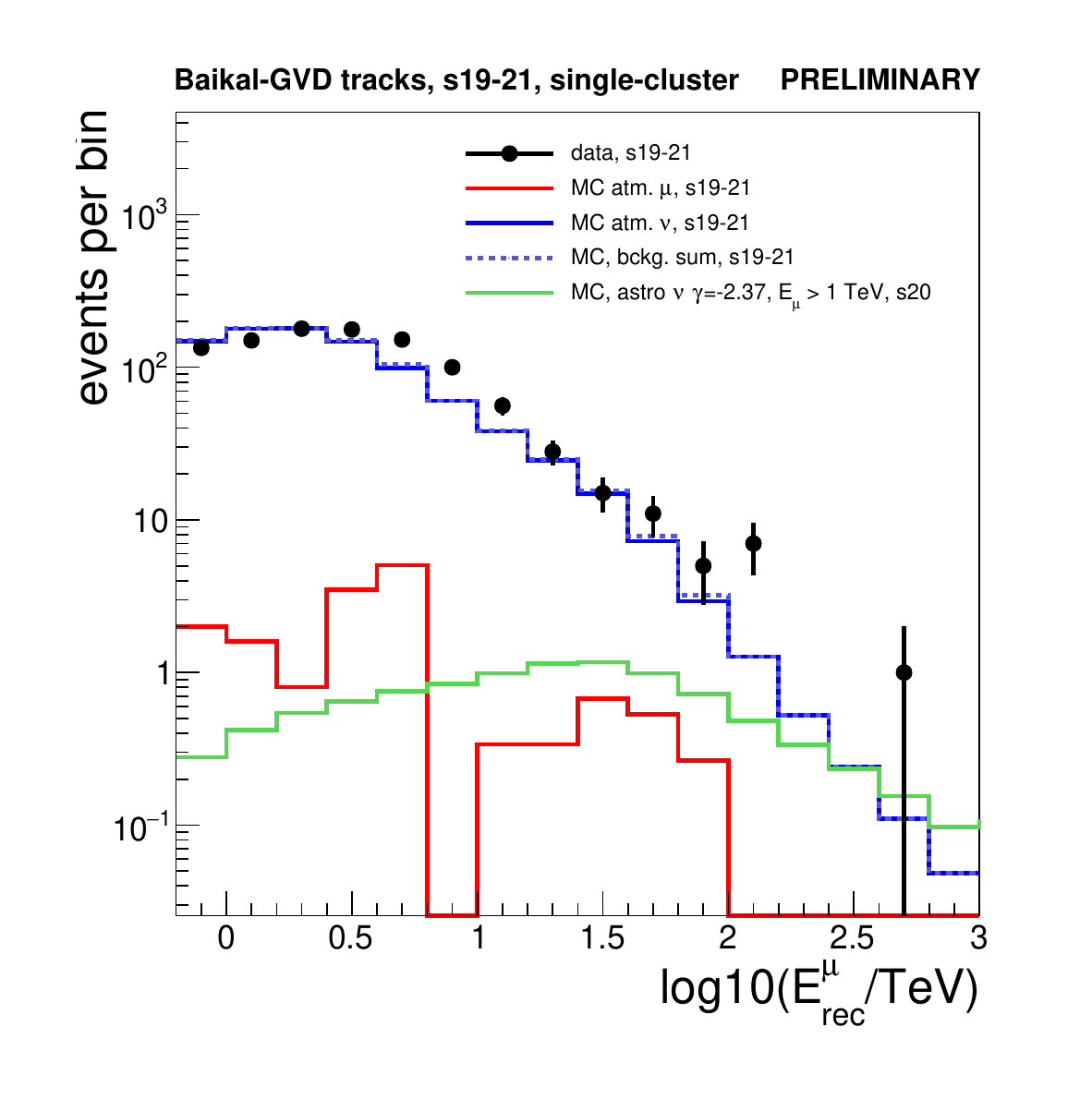}
	%\caption{}
	\end{subfigure}
	\caption{Distributions of number of track hits (a) and reconstructed muon energy (b) obtained in the data from seasons 2019-2021 compared to the expectations from MC.}
\end{center}
\end{figure}

\end{document}